\documentclass[twocolumn,showpacs,preprintnumbers,amsmath,amssymb]{revtex4}

\usepackage{graphicx}
\usepackage{dcolumn}
\usepackage{bm}

\begin{document}

\preprint{ }

\title{New Josephson Plasma Modes in Underdoped YBa$_2$Cu$_3$O$_{6.6}$ Induced by Parallel Magnetic Field}

\author{K. M. Kojima}\email{kojima@lyra.t.u-tokyo.ac.jp}
\author{S. Uchida}%
\affiliation{%
Graduate School of Frontier Sciences, University of Tokyo, 
Hongo 7-3-1, Bunkyo, Tokyo 113-8656, Japan
}%
\author{Y. Fudamoto}%
\author{S. Tajima}%
\affiliation{%
Superconductivity Research Laboratory, ISTEC,
Shinonome 1-10-13, Koto-ku, Tokyo 135-0062, Japan
}%

\date{ }

\begin{abstract}
The $c$-axis reflectivity spectrum of underdoped YBa$_2$Cu$_3$O$_{6.6}$
(YBCO) is measured below $T_c=59$K in parallel magnetic fields 
$H//$CuO$_2$ up to 7~T. Upon application of a parallel field, 
a new peak appears 
at finite frequency in the optical conductivity at the expense of suppression 
of $c$-axis condensate weight. We conclude that the dramatic change originates 
from different Josephson coupling strengths between bilayers with and without 
Josephson vortices. We find that the 400cm$^{-1}$ broad conductivity peak in 
YBCO gains the spectral weight under parallel magnetic field; this 
indicates that the condensate weight at $\omega =0$ is distributed to the 
intra-bilayer mode as well as to the new optical Josephson mode.
\end{abstract}

\pacs{74.25.Gz, 74.72.Bk, 74.50.+r}
\keywords{high Tc cuprates, Josephson plasma, parallel field}
\maketitle

Josephson plasma oscillation, which generally appears in the 
$E// c$ optical response of  high $T_c$ cuprates is a direct consequence
of the two-dimensionality of their superconductivity 
\cite{TamasakuPRL92,TachikiPRB94,BulaevskiiPRL96,OhashiJPSJ97}.
Depending on the anisotropy parameter $\gamma=\lambda_c/\lambda_{ab}$, 
the Josephson plasma frequency ranges from 
the microwave ($\gamma\sim 300-1000$ for
Bi$_2$Sr$_2$CaCu$_2$O$_{8+\delta}$ (Bi-2212) 
\cite{MatsudaPRL95,KadowakiPRL97,MotohashiPRB00}) 
to the far infrared frequency ($\gamma\sim 50$ for underdoped YBCO 
\cite{HomesPRL93,HauffPRL96,TajimaPRB97,BernhardPRB00,FukuzumiPRB00} 
and La$_{2-x}$Sr$_x$CuO$_4$ (LSCO) \cite{TamasakuPRL92,UchidaPRB96}). 
The optical/microwave response originates from the
restoration of coherent transport due to the tunneling of superconducting
carriers between the CuO$_2$ layers. The tunneling current, which is
driven by the phase-shift between the CuO$_2$ layers, is most directly
perturbed by the magnetic field applied parallel to the layers; the
interplay between parallel field and Josephson tunneling has recently 
become an interesting subject. Since the tunneling current is involved in the
screening of the parallel field, vortices have a large size of $\sim\gamma s$ 
within the layers, whereas in the $c$-axis direction,
they are squeezed to the separation $s$ of the superconducting layers.
This very anisotropic vortex in the parallel field is called a Josephson 
vortex.

In Bi-2212, there have been microwave resonance measurements under 
parallel applied magnetic fields 
\cite{KadowakiPHYSC01,KakeyaCondmat0111094,MatsudaPRB97},
 and resonance peaks have been 
observed in microwave absorption spectra. However, in the microwave study,
it is not possible to discuss the origin of the observed modes 
quantitatively. In this Letter, we report a dramatic change of 
the Josephson
plasma mode and an emergence of new modes under parallel fields in 
underdoped YBCO, where relatively small anisotropy $\gamma\sim 50$ 
brings Josephson resonance frequencies 
up to the far infrared regime. The technique of optical reflectivity 
was employed which enabled us to investigate both the resonance
energies of longitudinal and transverse modes {\it and } their 
oscillator strengths, necessary for a quantitative discussion
of the sum rule.

Another key issue of the $c$-axis response of YBCO is the effect of 
the bi-layer structure. Observations of a broad peak at $\sim 400$cm$^{-1}$ 
in the optical conductivity which appears below the pseudo-gap temperature 
have been made by several groups \cite{HomesPRL93,HauffPRL96,TajimaPRB97,BernhardPRB00,FukuzumiPRB00}. 
This ``400cm$^{-1}$'' peak
has been interpreted as the out-of-phase coupling of the two types of 
Josephson junctions in YBCO, which correspond to the intra- and the 
inter-bilayer coupling \cite{MunzarSSC99,GruningerPRL00}. 
Similar observations of this ``2nd 
Josephson plasma mode'' have been made in Bi-2212 \cite{ZeleznyPRB01} with
the same double pyramid structure as YBCO and $T^*$-type compounds, such as 
SmLa$_{1-x}$Sr$_x$CuO$_4$ \cite{ShibataPRL01,KakeshitaPRL01,DulicPRL01a},
which also has two types of insulating layers. 
Since Josephson vortices penetrate more easily into junctions with
weaker Josephson coupling, the applied field is expected to 
primarily modulate the inter-bilayer coupling in YBCO. It is also interesting 
to explore the effect of parallel field on the 400cm$^{-1}$ mode. 
If there were any effect, it would be direct evidence that 
superconductivity is involved in the formation of this mode. 

The specimen we measured is a YBa$_2$Cu$_3$O$_{6.6}$ single crystal
grown by the pulling method \cite{Shiohara97}. A large
crystal ($10\times 10\times 10$ mm$^3$) was cut along the $a\times c$-plane, 
post annealed in 1atm O$_2$ gas at 750$^\circ$C
in order to ensure the oxygen content, and then quenched to the room
temperature. Magnetic susceptibility (not shown) exhibits a sharp 
superconducting transition at $T_c=59$K, with the transition width of 
less than 2K. 
The crystal was polished using Al$_2$O$_3$ powders, and varnished on 
a sample holder with two $\phi 6$ mm holes, together with a 
gold coated glass mirror.
We employed a Fourier-transform infrared spectrometer
IFS-113v (Bruker) which is equipped with a superconducting magnet 
SpectroMag (Oxford). We performed calibration measurements 
in a regular cryostat in zero-field. The result agreed 
with that measured in the magnet. Small misalignments of the magnetic
field with respect to the CuO$_2$ plane direction do not have serious 
effects on the result, as the anisotropy is much smaller than that of Bi-2212. 

\begin{figure}
\includegraphics[width=\columnwidth]{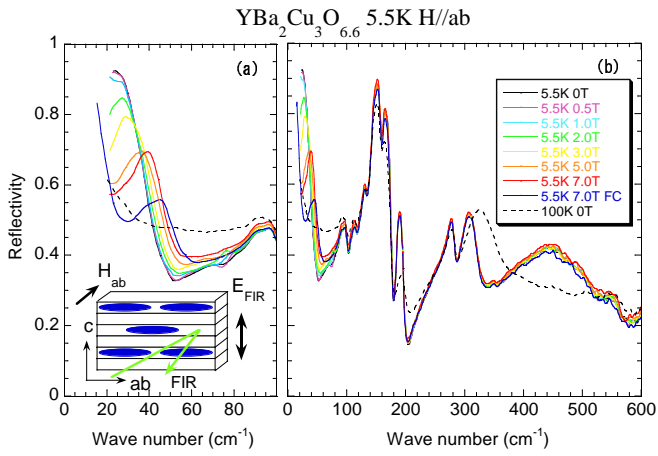}
\caption{\label{fig:fig1} $c$-axis reflectivity of YBCO, under an
applied magnetic field $H//ab$-plane. All the data shown are taken
after zero-field cooling, except the data labeled 7T FC 
(field cooling). }
\end{figure}

In Figs.\ref{fig:fig1}(a) and \ref{fig:fig1}(b), we show the $c$-axis optical 
reflectivity spectra of YBCO at 5.5K and various fields. At zero magnetic 
field, the Josephson plasma edge is located at $\sim 45$cm$^{-1}$ and a broad
bump at $\sim 400$cm$^{-1}$ is observed as has been previously reported 
\cite{HomesPRL93,HauffPRL96,TajimaPRB97,BernhardPRB00,FukuzumiPRB00}. 
When a parallel magnetic field ($H_{ab}\geq 1$ Tesla) is applied after the 
sample has been cooled to 5.5K in zero-field (zero-field cooling: ZFC), the  
reflectivity edge shifts to higher energy, but simultaneously the 
reflectivity in the low energy side rapidly decreases, forming a peak.
This result demonstrates a surprising sensitivity to the parallel magnetic 
field. By contrast, a slight and monotonic low energy shift is observed 
for fields applied perpendicular to the planes (not shown). 
Fig.\ref{fig:fig1} also shows the data taken after field-cooling at 
7T (labeled as ``7.0T FC''). In this field-cooling situation, the reflectivity
increases below $\sim 25$cm$^{-1}$ forming an additional edge, and the 
location of the higher energy peak and edge structure is different from 
that at 7T(ZFC). The spectral difference between FC and ZFC indicates that the 
observed spectral change is related to the pinning of Josephson vortices.

\begin{figure}
\includegraphics[width=\columnwidth]{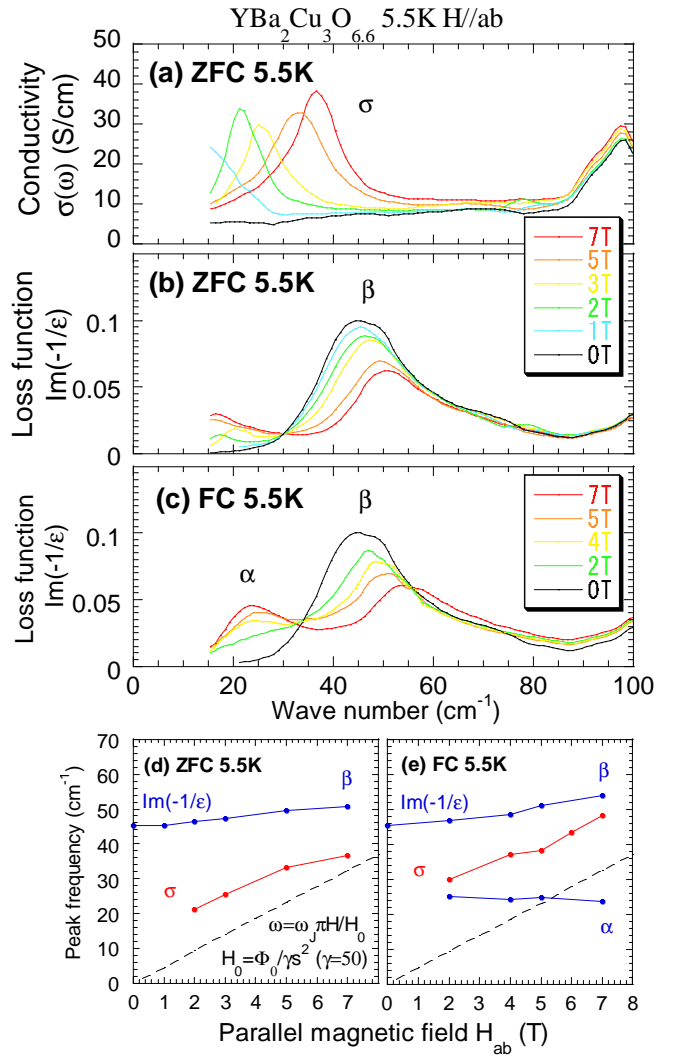}
\caption{\label{fig:fig2} Optical conductivity (a) and energy
loss function (b) at 5.5K after zero-field cooling. 
(c) Energy loss function at 5.5K after field cooling.
Measured peak position of (d) ZFC and (e) FC. 
The dashed line in (d) and (e) is a theoretical prediction for the
high-field limit based on Ref. \cite{BulaevskiiPRB97}.}
\end{figure}

To make a quantitative analysis of the data, we employed 
the Kramers-Kronig analysis of the reflectivity, and calculated the 
optical conductivity $\sigma(\omega)=\omega Im(\varepsilon)/4\pi$ and 
the energy loss function $Im(-1/\varepsilon)$. The results are shown 
in Figs.\ref{fig:fig2}(a)-(c). 
A remarkable change in the spectrum is the appearance of a peak 
in the optical conductivity, labeled as ``$\sigma$'' in 
Fig.\ref{fig:fig2}(a), which shifts to higher energies with increasing field.
The peak intensity, on the other hand, does not depend
much on the field strength. The $\sigma$-mode obtains appreciable intensity 
already at $H_{ab}=2$T (probably at an even lower field in view of the 
data at 1T in Fig.\ref{fig:fig2}(a)), which suggests an abrupt growth of
intensity upon application of a parallel field. 
In Figs.\ref{fig:fig2}(b) and \ref{fig:fig2}(c), the 
energy-loss functions 
which generally measure longitudinal modes are compared for zero-field 
cooling (ZFC) and field-cooling (FC). In ZFC [Fig.\ref{fig:fig2}(b)], 
there is only one peak observed in our experimental window 
($\omega>15$cm$^{-1}$).  
In FC [Fig.\ref{fig:fig2}(c)], there are two peaks emerge within our window.
The new longitudinal mode (labeled  ``$\alpha$'') increases 
its intensity with $H_{ab}$. The original longitudinal mode (labeled 
 ``$\beta$'', corresponding to the Josephson plasma mode at $H_{ab}=0$) 
decreases in intensity, showing that the spectral weight transferred
from $\beta$- to $\alpha$-mode. This happens also for ZFC, though the 
$\alpha$-peak is not seen above 15cm$^{-1}$. The energy positions of the 
peaks are plotted in Figs.\ref{fig:fig2}(d) and \ref{fig:fig2}(e) as a 
function of $H_{ab}$. The 
peak energy of the $\beta$-mode shifts upward, whereas the
peak energy of the $\alpha$-mode decreases slightly with $H_{ab}$. 

\begin{figure}
\includegraphics[width=\columnwidth]{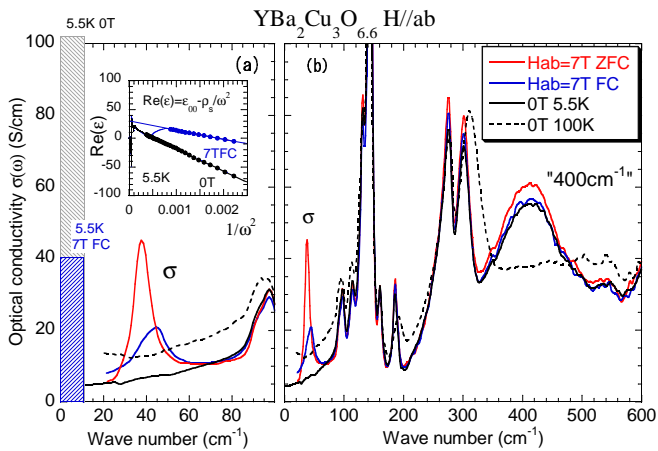}
\caption{\label{fig:fig3} (a) $c$-axis optical conductivity of YBCO, under a
parallel field $H_{ab}$. The rectangles indicate
the weight of low-energy condensate as derived from fitting to the
function $Re(\varepsilon) \approx \varepsilon_\infty-\rho_s/\omega^2$ 
shown in the inset for 0T and 7T(FC) at 5.5K. The filled circles in the 
inset indicate the data points used for the fitting.
(b) Optical conductivity on a wider scale.}
\end{figure}

Existence of longitudinal $\alpha$-mode, or equivalently, the reflectivity 
edge below the peak energy, indicates that there is another transverse
mode at $\omega_{\sigma'}<15$cm$^{-1}$. This $\sigma'$-mode
may be either an $\omega=0$ superconducting condensate or a dissipative mode
at finite energy.
We estimated the spectral weight of this $\sigma'$-mode ($\rho_s$) by 
analyzing the low-energy asymptotic behavior of the dielectric 
function $Re(\varepsilon)\approx \varepsilon_\infty-\rho_s/\omega^2$ 
(see inset of Fig.\ref{fig:fig3}(a)). This analysis is exact for
superconducting condensate, but may have $\sim \pm30$\% error in the spectral 
weight if $\omega_{\sigma'}$ is finite.
For $H_{ab}=0$ $\rho_s$ is a genuine $c$-axis condensate weight, and 
estimated to be $\rho_s=4.0(1)\times 10^4$ cm$^{-2}$. For $H_{ab}=7$T(FC), 
there remains a weight of $1.5(4)\times 10^4$cm$^{-2}$ at low energy
which locates either at $\omega=0$ or at a finite energy $\omega_{\sigma'}$.
The low energy condensate weights are shown in Fig.\ref{fig:fig3}(a) 
as the height of the rectangles with a width of 10cm$^{-1}$.
For measurements at 7T(ZFC), $1/\omega^2$ behavior in $Re(\varepsilon)$ 
is no longer seen, which suggests that the low-$\omega$ weight is very small. 

Optical conductivity over wider energy range is also shown in 
Fig.\ref{fig:fig3}(b). It is clear that the conductivity 
peak at 400cm$^{-1}$ gains intensity under magnetic fields, 
especially at 7T(ZFC). 
The integrated intensity of the $\sigma$-peak and the 
increase in the 400cm$^{-1}$ peak intensity yield
$S(\omega_\sigma)=1.8(1) \times 10^4$ cm$^{-2}$ and 
$\Delta S(\omega_T)=2.2(2) \times 10^4$ cm$^{-2}$ respectively at 7T(ZFC), 
which completely exhausts the weight of superconducting condensate 
at zero-field.
On the other hand, at 7T(FC), $S(\omega_\sigma)=1.1(1)\times 10^4$cm$^{-2}$ 
and $\Delta S(\omega_T)=0.5(2)\times 10^4$cm$^{-2}$. The sum explains 60-70\%
of the missing $\rho_s$ ($\sim 2.5(4)\times 10^{4}$cm$^{-1}$) transferred
to the $\sigma$-mode and the 400cm$^{-1}$ peak. Substantial weight remains 
in the $\sigma'$-mode, which is also evidenced by the evolution of 
the $\alpha$-peak in $Im(-1/\varepsilon)$ with increasing $H_{ab}$ seen
in Fig. \ref{fig:fig2}(c). 
The sum rule analysis gives evidence for spectral-weight transfer from
the superconducting condensate to the other two or three finite-$\omega$ 
transverse modes, enforced by parallel fields or by modulation of 
the inter-/intra-bilayer coupling.

Now we discuss the origin of the $\sigma$-mode. It is proposed
that Josephson vortex lattice has a quantized $c$-axis lattice parameter 
($=N\times c$), whereas its in-plane lattice parameter is tunable 
to match the external field \cite{BulaevskiiPRB91}. 
In this situation, the inter-bilayer coupling is expected to be modulated, 
differentiating the Josephson coupling strength between bilayers with and 
without Josephson vortices, giving rise to a transverse (optical) Josephson 
plasma mode at 
finite $\omega$. For underdoped YBCO, the anisotropy 
parameter $\gamma\sim 50$ scales the characteristic field 
$H_0=\Phi_0/\gamma s^2$ to $\approx 30$T \cite{BulaevskiiPRB91,IchiokaPRB95}.
Calculations of free-energies show that Josephson vortices exist in one out 
of every two or more layers when $H_{ab}\lesssim H^*=H_0/3 
\approx 10$T \cite{BulaevskiiPRB91,IchiokaPRB95}, 
which very nearly matches our experimental conditions. 
A phenomenological model which describes the optical response
of a Josephson junction array with more than two kinds of insulating 
layers has been proposed by van der Marel {\it et al.} \cite{MarelCJP96}
which is called as ``multi-layer model''. This model has been employed
for the analysis of $c$-axis optical response of bi-layer compounds 
in zero-field (YBCO \cite{MunzarSSC99,GruningerPRL00} and 
Bi-2212 \cite{ZeleznyPRB01})
and $T^*$ materials \cite{ShibataPRL01,KakeshitaPRL01,DulicPRL01a} 
where two kinds of Josephson couplings are expected due to the crystal 
structure. We model YBCO in parallel magnetic fields by a simplified 
multi-layer model which takes into account only the low energy modes
in the energy range ($\omega<80$cm$^{-1}$) where optical phonons are absent.
We assume that the possible $\sigma'$-mode is located at $\omega=0$: 
\begin{eqnarray}
\label{eq:eps-3layer}
\frac{\varepsilon(\omega)}{\varepsilon_\infty} &=& 
\frac{(\omega^2-\omega_\alpha^2)(\omega^2-\omega_\beta^2)}
     {\omega^2(\omega^2-\omega_\sigma^2)}
\end{eqnarray}
After a simple calculation, the spectral weight of the transverse modes 
$S(\omega_\sigma$) and $\rho_s$ are given by a function of the energies 
of these resonance modes. At $H_{ab}=7$T(FC), the estimate is
straightforward, since all the relevant resonance frequencies are observed. 
They are $\omega_\alpha=23.6$cm$^{-1}$, $\omega_\sigma=44.4$cm$^{-1}$ and 
$\omega_\beta=54.0$cm$^{-1}$. Using these frequencies and the dielectric 
constant $\varepsilon_\infty=21$, the calculation gives 
$S(\omega_\sigma)=1.4\times 10^{4}$cm$^{-2}$ and 
$\rho_s=1.7\times 10^{4}$cm$^{-2}$, in reasonable agreement with the 
observed $\sigma$-peak intensity 
$S(\omega_\sigma)=1.1(1)\times 10^4$cm$^{-2}$ and
the $\sigma'$-peak intensity (or superconducting condensate)
$\rho_s=1.5(4)\times 10^4$cm$^{-2}$. At 7T(ZFC), the location of 
the $\alpha$-peak is uncertain which makes the estimate somewhat ambiguous. 
Also in this case, the observed $\sigma$-peak intensity 
$1.8(1)\times 10^{4}$cm$^{-2}$
is consistent with a rough estimate of $\approx 2\times 10^{4}$cm$^{-2}$.
These agreements supports the hypothesis that the $\sigma$-mode 
originates from the periodic modulation of inter-bilayer Josephson 
coupling strength due to the Josephson vortices. 

A theoretical investigation of Josephson plasma in parallel field was
performed by Bulaevskii {\it et al.} under the ideal condition of 
absence of pinning to the Josephson vortices \cite{BulaevskiiPRB97}.
They found two strong resonance modes: one corresponding to
``vortex sliding mode'', and the other arising from the Josephson 
plasma dispersion folded to the zone center by the formation of 
a vortex superlattice \cite{BulaevskiiPRB97}.
The spectral weight is transferred from the latter to the former 
with $H_{ab}$.
In real materials, the vortex sliding mode is expected to be influenced by 
pinning of the Josephson vortices; the $\sigma'$-mode (or the associated 
longitudinal $\alpha$-mode) in YBCO depends on the field-cycle and may 
be assigned to the pinned sliding mode of Josephson vortices.
The energy of the ``folded Josephson plasma mode'' calculated 
in high field limit \cite{BulaevskiiPRB97}, presents a linear field dependence 
as shown in Figs.\ref{fig:fig2}(d) and \ref{fig:fig2}(e) by the 
dashed line. 
It appears that the energy of the $\sigma$-mode (or the associated 
$\beta$-mode) in YBCO tends to approach this line at higher fields. In fact,
the spectral weight transfer between $\alpha$- and $\beta$-mode shown in Figs.
\ref{fig:fig2}(b) and \ref{fig:fig2}(c) is consistent with the theoretical
result \cite{BulaevskiiPRB97}. Recent microwave measurement of Bi-2212 
under parallel field has 
revealed two resonance modes \cite{KakeyaCondmat0111094}. In the light of the 
present results for YBCO, it is likely that the higher energy 
mode corresponds to the $\beta$- (or $\sigma$-) mode, and the low-energy mode 
to the $\alpha$-mode (or $\sigma'$-mode). 

The reason why the transverse $\sigma$- and 400cm$^{-1}$ modes are 
considerably more enhanced in ZFC than in FC might be
related to the pinning of Josephson vortices. 
As the alternating stack of insulating layers with and without Josephson 
vortices is most likely the source of the $\sigma$-mode, disorder in 
stacking would reduce the $\sigma$-mode intensity. This seems to be the case
with the $\sigma$-mode for FC, since random distribution of pinning centers 
tends to make the stacking of Josephson vortices more disordered. 
In a $T^*$ cuprate Nd$_{2-x-y}$Sr$_x$Ce$_y$CuO$_4$, 
reduced intensity of the transverse Josephson plasma mode has been 
interpreted as the result of disorder in the crystalline structure 
\cite{KakeshitaPRL01}. 

In LSCO, no new transverse $\sigma$-mode under parallel field is seen 
\cite{KimuraJPSJ96,GerritsPRB95,BentumPHYSC97}, 
although the anisotropy parameter $\gamma$ is in the same order of 
magnitude with underdoped YBCO. Only a low-$\omega$ shift of the 
Josephson plasma edge is observed with increasing parallel field, 
suggesting that the Josephson vortices would be highly disordered
seemingly due to fluctuations of stripe order in LSCO. 

In conclusion, we have demonstrated a remarkable effect of parallel 
magnetic field on the $c$-axis Josephson plasma mode in underdoped YBCO. 
Parallel magnetic field modulates the inter-bilayer Josephson coupling strength
along the $c$-axis due to the presence of inequivalent insulating layers 
with and without 
Josephson vortices. As a result, one optical (transverse) mode appears,
corresponding to the anti-phase Josephson current oscillations between two
inequivalent junctions. Another consequence of the modulation of the 
inter-bilayer coupling and/or suppression of inter-bilayer coherent pair 
hopping is the enhanced intensity of the 400cm$^{-1}$ mode. This gives 
evidence that this mode is a marker of superconducting Josephson coupling 
within a bilayer. 

The authors would like to thank Prof. M.~Tachiki, Drs. Y.~Ando,
M.~Machida, I.~Kakeya and T.~Kakeshita for stimulating discussions.
The research in this paper has been financially supported
by NEDO International Joint Research Grant and by
COE \& Grant-in-aid for Scientific Research from Monbusho.

\end{document}